# Complementarity problems for two pairs of charged bodies

A.A. Kolpakov (Université de Fribourg, Suisse),

A.G. Kolpakov (Marie Curie Fellow, Novosibirsk, Russia)

We consider an interaction of charged bodies under the following simplified conditions: the distribution of charge over each body is stable; the interaction of bodies is governed by electrical forces only.

Physically, these assumptions can be treated as the following decomposition of charges: the structure of each body is assumed to be stable due to inner forces (say, quantum forces [1]), which do not influence the interaction of the bodies; the bodies interact due to the classical electrical forces [2] only. In this model, the role of inner forces is to create a specific stable distribution of the charge over a body.

We assume that the charge distribution over a body can be described by the density of the charge. In our model, the distribution of the charge is the property of a body and does not change in the process of the bodies' interaction. For the simplicity we assume that the bodies are similar in the sense of geometry, say, occupy domain $Q$ and have a preferable direction of interaction denoted by $Ox_3$.

**Statement of the complementarity problem.** The problem we are considering is the complementation problem formulated as follows: Do there exist a charge distribution that makes the bodies to form pairs, such that the bodies in each pair attract one to another while push away other bodies?

For one pair of distributions, the problem has a trivial solution: $\phi(\mathbf{x}) = 1$ and $\Phi(\mathbf{x}) = -1$. The first nontrivial case is the case of two pair of distributions.

We consider the case when two bodies are placed one in front of the other (like two bodies on the left-hand side of Fig.1). In this case, the interaction of two charges with distributions $\phi(\mathbf{x})$ and $\psi(\mathbf{y})$ is measured by the interaction force in direction $Ox_3$

$$\mathbf{e}_3 \mathbf{F} = \mathbf{e}_3 \mathbf{F}(|\mathbf{x}-\mathbf{y}|)\phi(\mathbf{x})d\mathbf{x}\psi(\mathbf{y})d\mathbf{y} \tag{1}$$

Denote

$$F(|\mathbf{x}-\mathbf{y}|) = \mathbf{e}_3 \mathbf{F}(|\mathbf{x}-\mathbf{y}|) \tag{2}$$

The total force for given distributions – pair interaction force

$$I\langle\phi,\psi\rangle = \int_Q \int_{Q+\mathbf{d}} F(|\mathbf{x}-\mathbf{y}|)\phi(\mathbf{x})d\mathbf{x}\psi(\mathbf{y})d\mathbf{y} = \int_Q \int_Q F(|\mathbf{x}-\mathbf{d}-\mathbf{y}|)\phi(\mathbf{x})d\mathbf{x}\psi(\mathbf{y})d\mathbf{y} \tag{3}$$

The vector $\mathbf{d}$ is displayed in Fig.1, the domains $Q$ and $Q+\mathbf{d}$ are two domains at the left-hand side in Fig.1.

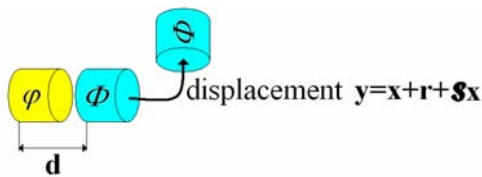

**Fig.1.** *Two bodies one in the front of other and the right body displaced*

Introduce operator

$$\mathsf{R}: \psi(\mathbf{y}) \in L_2(Q) \to \int_Q R(|\mathbf{x}-\mathbf{y}|)\psi(\mathbf{y})d\mathbf{y} \in L_2(Q), \qquad (4)$$

where $R(|\mathbf{x}-\mathbf{y}|) = F(|\mathbf{x}-\mathbf{d}-\mathbf{y}|)$.

With this notation, we have the following formula for the pair interaction force

$$I\langle \phi, \psi \rangle = (\phi, \mathsf{R}\psi), \qquad (5)$$

$(\,,\,)$ means the standard scalar product in $L_2(Q)$.

**The general case.** Generally speaking, every body has several degrees of freedom and moves in space. A movement of a body can be described by using the transformation $\mathbf{y} \to \mathbf{y} + \mathbf{r} + \mathsf{S}\mathbf{y}$, where $\mathbf{r}$ is a translation vector, $\mathsf{S}$ means operation of rotation. Generally, one has to change (3)-(5) for

$$I(\mathbf{r},\mathsf{S})\langle \phi, \psi \rangle = \int_Q \int_Q R(|\mathbf{x}-\mathbf{r}-\mathsf{S}\mathbf{y}|)\phi(\mathbf{x})d\mathbf{x}\psi(\mathbf{y})d\mathbf{y} \qquad (6)$$

$$\mathsf{R}(\mathbf{r},\mathsf{S}): \psi(\mathbf{y}) \in L_2(Q) \to \int_Q R(|\mathbf{x}-\mathbf{r}-\mathsf{S}\mathbf{y}|)\psi(\mathbf{y})d\mathbf{y} \in L_2(Q) \qquad (7)$$

$$I(\mathbf{r},\mathsf{S})\langle \phi, \psi \rangle = (\phi, \mathsf{R}(\mathbf{r},\mathsf{S})\psi) \qquad (8)$$

We assume that contact position of a pair of bodies is $x_3 = 0$. We say two bodies with distributions $\phi$ and $\psi$ to be complementary if the pair interaction force $(\phi, \mathsf{R}(\mathbf{r},\mathsf{S})\psi) > 0$ for some $\mathbf{r}, \mathsf{S}$ and $(\phi, \mathsf{R}(\mathbf{r},\mathsf{S})\psi) < 0$ any $\mathbf{r}, \mathsf{S}$ in the contact position.

**A special case.** Let us consider two bodies with axisymmetrical distribution of charges as shown in Fig.2. We restrict the translation degrees of freedom in $Ox_1$ and $Ox_2$ directions and allow translation along $Ox_3$-axis, only. In this case, the limit position is $\mathbf{r} = 0$, Rotation is possible about the $Ox_3$-axis only and transforms $\mathbf{y} \to \mathsf{S}\mathbf{y}$ do not change the distribution of axisymmetrical charges. In the limit position, operator $\mathcal{R}$ is

$$\mathcal{R}\psi = \int_Q R(|\mathbf{x}-\mathbf{y}|)\psi(\mathbf{y})d\mathbf{y} \qquad (9)$$

The kernel $R(|\mathbf{x}-\mathbf{y}|)$ of the integral operator in (9) is the fundamental solution of the electrostatic problem and is singular [2]. We make a simplification and assume that the kernel is a smooth function (thus we assume that the charges belonging to one body do not touch the charges belonging to another body).

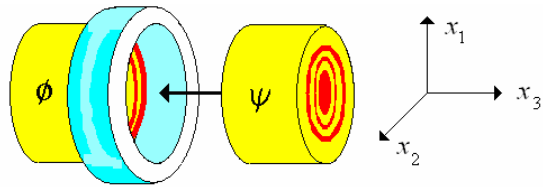

**Fig.2**. *A pair of bodies with axi-simmetrical distribution of charges.*

The linear integral operator $\mathcal{R}: L_2(Q) \to L_2(Q)$ (9) is compact, self-adjoin:

$$(\mathsf{R}\psi,\phi) = \int_Q \int_Q R(|\mathbf{x}-\mathbf{y}|)\psi(\mathbf{y})\phi(\mathbf{x})d\mathbf{y}d\mathbf{x} = (\mathsf{R}\phi,\psi), \quad (10)$$

and negatively determined

$$(\mathsf{R}\psi,\psi) = \int_Q \int_Q R(|\mathbf{x}-\mathbf{y}|)\psi(\mathbf{y})\psi(\mathbf{x})d\mathbf{y}d\mathbf{x} < 0. \quad (11)$$

**The problem of complementarity - Strong formulation.** Do exist four distributions $\phi$, $\psi$, $\Phi$ and $\Psi$ possessing the properties:
$\phi$ is complementary to $\Phi$ and not complementary to $\phi$, $\psi$ and $\Psi$;
$\psi$ is complementary to $\Psi$ and not complementary to $\phi$, $\psi$ and $\Phi$; also $\Phi$ is not complementary to $\Psi$.

Due to (5), the conditions above may be written in the form

$$I\langle\phi,\Phi\rangle > 0,\ I\langle\phi,\phi\rangle < 0,\ I\langle\phi,\psi\rangle < 0,\ I\langle\phi,\Psi\rangle < 0,\ I\langle\Phi,\Phi\rangle < 0,\ I\langle\Phi,\Psi\rangle < 0; \quad (12)$$

$$I\langle\psi,\Psi\rangle > 0,\ I\langle\psi,\psi\rangle < 0,\ I\langle\psi,\phi\rangle < 0,\ I\langle\psi,\Phi\rangle < 0,\ I\langle\Psi,\Psi\rangle < 0,\ I\langle\Phi,\Psi\rangle < 0.$$

**The problem of complementarity - Week formulation.** Does exist four distributions possessing the properties

$$I\langle\phi,\Phi\rangle > 0,\ I\langle\phi,\phi\rangle \le 0,\ I\langle\phi,\psi\rangle \le 0,\ I\langle\phi,\Psi\rangle \le 0,\ I\langle\Phi,\Phi\rangle \le 0,\ I\langle\Phi,\Psi\rangle \le 0; \quad (13)$$

$$I\langle\psi,\Psi\rangle > 0,\ I\langle\psi,\psi\rangle \le 0,\ I\langle\psi,\phi\rangle \le 0,\ I\langle\psi,\Phi\rangle \le 0,\ I\langle\Psi,\Psi\rangle \le 0,\ I\langle\Phi,\Psi\rangle \le 0.$$

**Solution for week formulation.** Linear operator $\mathcal{R}: L_2(Q) \to L_2(Q)$, which is compact and self-adjoint, has a system of eigenfunctions $\{\varphi_i\}_{i=1}^{\infty}$, that form an orthogonal basis in $L_2(Q)$ [3]. In particular,

$$\mathcal{R}\varphi_i = \lambda_i\varphi_i,\ (\varphi_i,\varphi_j) = \delta_{ij}. \quad (14)$$

We set

$$\phi = \varphi_i,\ \Phi = -\varphi_i,\ \psi = \varphi_j,\ \Psi = -\varphi_j,\ i \ne j. \quad (15)$$

Since $\mathcal{R}$ is negatively determined

$$I\langle\phi,\Phi\rangle > 0,\ I\langle\phi,\phi\rangle < 0,\ I\langle\psi,\Psi\rangle > 0,\ I\langle\psi,\psi\rangle < 0,\ I\langle\Phi,\Phi\rangle < 0,\ I\langle\Psi,\Psi\rangle < 0. \quad (16)$$

Since $\varphi_i$ and $\varphi_j$ are orthogonal eigenfunctions

$$I\langle\phi,\psi\rangle = 0,\ I\langle\phi,\Psi\rangle = 0,\ I\langle\psi,\phi\rangle = 0,\ I\langle\psi,\Phi\rangle = 0. \quad (17)$$

By virtue of (16), (17), the functions (15) satisfy all conditions in (13).

**Solution for strong formulation.** We construct a perturbation [4] of solution (15). Let $\phi$ and $\psi$ are solutions of the problem in the week formulation presented below. We take an eigenfunction $\varphi$ differ from $\phi$ and $\psi$, and form the following four linear combinations (see Table 1) where $\alpha$ is a small real number.

| | |
|---|---|
| $\tilde{\phi} = \phi + \alpha\varphi$ | $\tilde{\Phi} = \Phi + \alpha\varphi$ |
| $\tilde{\psi} = \psi + \alpha\varphi$ | $\tilde{\Psi} = \Psi + \alpha\varphi$ |

**Table 1.** Four linear combinations

By virtue of (16), we have for sufficiently small $\alpha$

$$I\langle\tilde{\varphi},\tilde{\Phi}\rangle > 0,\ I\langle\tilde{\varphi},\tilde{\varphi}\rangle < 0,\ I\langle\tilde{\psi},\tilde{\Psi}\rangle > 0,\ I\langle\tilde{\psi},\tilde{\psi}\rangle < 0,\ I\langle\tilde{\Phi},\tilde{\Phi}\rangle < 0,\ I\langle\tilde{\Psi},\tilde{\Psi}\rangle < 0, \quad (18)$$

and, by virtue of (14),

$$I\langle\tilde{\phi},\tilde{\psi}\rangle = (\tilde{\phi},\mathcal{R}\tilde{\psi}) = (\phi+\alpha\varphi,\mathcal{R}\psi+\alpha\mathcal{R}\varphi) = (\phi+\alpha\varphi,\lambda_\psi\psi+\alpha\lambda_\varphi\varphi) = \alpha^2\lambda_\varphi\|\varphi\|^2, \quad (19)$$

$$I\langle\tilde{\phi},\tilde{\Psi}\rangle = (\tilde{\phi},\mathcal{R}\tilde{\Psi}) = (\phi+\alpha\varphi,-\mathcal{R}\psi+\alpha\mathcal{R}\varphi) = (\phi+\alpha\varphi,-\lambda_\psi\psi+\alpha\lambda_\phi\varphi) = \alpha^2\lambda_\phi\|\varphi\|^2,$$

$$I\langle\tilde{\psi},\tilde{\phi}\rangle = I\langle\tilde{\phi},\tilde{\psi}\rangle = \alpha^2\lambda_\varphi\|\varphi\|^2,$$

$$I\langle\tilde{\psi},\tilde{\Phi}\rangle = (\tilde{\psi},\mathcal{R}\tilde{\Phi}) = (\psi-\alpha\varphi,-\mathcal{R}\varphi+\alpha\mathcal{R}\varphi) = (\phi+\alpha\varphi,-\lambda_\phi\phi+\alpha\lambda_\varphi\varphi) = \alpha^2\lambda_\varphi\|\varphi\|^2.$$

Here $\lambda_\phi$, $\lambda_\psi$ and $\lambda_\varphi$ are the eigenvalues corresponding to the eigen-functions $\phi$, $\psi$ and $\varphi$, correspondingly, $\|\ \|$ means the standard norm in $L_2(Q)$. Since the operator $\mathcal{R}$ is negatively determined, those eigenvalues are negative. As a result, $\alpha^2\lambda_\varphi\|\varphi\|^2 < 0$.

Deriving (19), we have used that the eigenfunctions $\phi$, $\psi$ and $\varphi$ are orthogonal.

**Conclusions.** Since just a simple model has been considered, all our conclusions are preliminary. The splitting of the interaction problem for charged bodies into two, namely
1. the problem of quantum mechanics, responsible for the stability of the bodies,
2. the electrostatic problem, responsible for the interaction of the bodies
is an effective simplification method for the initial problem.

In the framework of this approach, the problem of complementarity can be solved, at least in a simple modeling case. It is seen that the problem of complementary is a so called existence problem, a problem about the existence of functions satisfied a system of inequalities.

The solutions obtained (say again, obtained in the frameworks of simple model) demonstrates that from the mathematical point of view there exist a great variety of complementary charged structures.

**Acknowledgments.** AGK was supported through Marie Curie actions FP7: project PIIF2-GA-2008-219690.